\documentclass[a4paper,11pt]{article}

\usepackage{amsmath,amssymb,bbm}
\usepackage{ascmac}
\usepackage{cancel}
\usepackage{graphicx}
\usepackage[bookmarks=true,bookmarksnumbered=true,bookmarkstype=toc]{hyperref}
\hypersetup{
pdfauthor={Tetsuji Kimura, Shin Sasaki and Masaya Yata},
colorlinks={true},
linkcolor={blue},
urlcolor={blue},
citecolor={blue}
}
\usepackage{ulem}

\makeatletter

 \@addtoreset{equation}{section}
\makeatother

\newcounter{Enumerate}

\DeclareFontFamily{U}{rsf}{}
\DeclareFontShape{U}{rsf}{m}{n}{
  <5> <6> rsfs5 <7> <8> <9> rsfs7 <10-> rsfs10}{}
\DeclareMathAlphabet\Scr{U}{rsf}{m}{n}

\usepackage[mathscr]{eucal}

\newcommand{\half}{\frac{1}{2}}

\newcommand{\ls}{\ \ \ \ \ }
\newcommand{\wt}{\widetilde}
\newcommand{\wh}{\widehat}

\newcommand{\bsubeq}{\begin{subequations}}
\newcommand{\esubeq}{\end{subequations}}

\newcommand{\noi}{\noindent}

\newcommand{\I}{{\rm i}}
\newcommand{\N}{\mathcal{N}}
\newcommand{\T}{{\rm T}}
\renewcommand{\d}{{\rm d}}
\newcommand{\e}{{\rm e}}

\newcommand{\slb}{\scalebox}

\def\+{{+\!\!\!+}} 

\begin{document}
\allowdisplaybreaks{

\thispagestyle{empty}


\begin{flushright}
KEK-TH-1778,
TIT/HEP-641 
\end{flushright}

\vspace{30mm}

\noi
\slb{2.5}{Hyper-K\"{a}hler with Torsion, T-duality,}

\vspace{5mm}

\noi
\slb{2.5}{and Defect $(p,q)$ Five-branes}

\vspace{10mm}

\slb{1.2}{Tetsuji {\sc Kimura}$^{\,a}$}, \
\slb{1.2}{Shin {\sc Sasaki}$^{\,b}$} 
\ and \ 
\slb{1.2}{Masaya {\sc Yata}$^{\,c}$}

\slb{.85}{\renewcommand{\arraystretch}{1.2}
\begin{tabular}{rl}
$a$ & 
{\sl 
Department of Physics,
Tokyo Institute of Technology} 
\\
& {\sl 
Tokyo 152-8551, JAPAN
}
\\
& {\tt tetsuji \_at\_ th.phys.titech.ac.jp}
\end{tabular}
}

\slb{.85}{\renewcommand{\arraystretch}{1.2}
\begin{tabular}{rl}
$b$ & {\sl
Department of Physics,
Kitasato University}
\\
& {\sl 
Sagamihara 252-0373, JAPAN}
\\
& {\tt shin-s \_at\_ kitasato-u.ac.jp}
\end{tabular}
}

\slb{.85}{\renewcommand{\arraystretch}{1.2}
\begin{tabular}{rl}
$c$ & {\sl
KEK Theory Center, Institute of Particle and Nuclear Studies,}
\\
& {\sl 
High Energy Accelerator Research Organization (KEK)
}
\\
& {\sl 
Tsukuba, Ibaraki 305-0801, JAPAN}
\\
& {\tt yata \_at\_ post.kek.jp}
\end{tabular}
}


\vfill


\slb{1.1}{\sc Abstract}
\begin{center}
\slb{.95}{
\begin{minipage}{.95\textwidth}
We investigate a five-branes interpretation of hyper-K\"{a}hler geometry with torsion (HKT).
This geometry is obtained by conformal transformation of the Taub-NUT space which represents a Kaluza-Klein five-brane.
This HKT would represent an NS5-brane on the Taub-NUT space.
In order to explore the HKT further, 
we compactify one transverse direction, and study the $O(2,2;{\mathbb Z}) = SL(2,{\mathbb Z}) \times SL(2,{\mathbb Z})$ monodromy structure associated with two-torus.
Performing the conjugate transformation,
we obtain a new solution whose physical interpretation is a defect $(p,q)$ five-brane on the ALG space.
Throughout this analysis, 
we understand that the HKT represents a coexistent state of two kinds of five-branes.
This situation is different from composite states such as $(p,q)$ five-branes or $(p,q)$ seven-branes in type IIB theory.
We also study the T-dualized system of the HKT.
We again find a new solution which also indicates another defect $(p,q)$ five-brane on the ALG space. 
\end{minipage}
}
\end{center}


\newpage
\section{Introduction}
\label{sect:introduction}

It is well known that target spaces of supersymmetric nonlinear sigma models with four and eight supercharges are K\"{a}hler and hyper-K\"{a}hler geometries, respectively \cite{Zumino:1979et, AlvarezGaume:1981hm}.
These geometries do not involve three-form fluxes in supergravity solutions.
However, the mathematical analysis for manifolds with non-zero fluxes was considered later by Strominger \cite{Strominger:1986uh}. 
On a K\"{a}hler manifold, the metric is required to be Hermitian with respect to the complex structure, 
and the complex structure must be covariantly constant. 
In the case of a hyper-K\"{a}hler manifold, three complex structures obeying the quaternionic multiplication law require the same condition. 
When there exists a non-zero flux, 
the covariant derivative includes the non-zero flux as a torsion. 
The K\"{a}hler geometry with torsion (KT) structure is defined by the covariantly constant complex structure including the non-zero torsion, 
and the hyper-K\"{a}hler geometry with torsion (HKT) structure is defined similarly. 
When the three-form flux $H$ is closed ($\d H=0$), the structure is called {\sl strong}, while if not it is called the {\sl weak} structure. 
The connections in these geometries are called the KT- and HKT-connections. 
Mathematicians often call the connections the Bismut connections \cite{Mavromatos:1987ru, Bismut:1989}. 
Manifolds with non-zero torsion have been studied in various contexts.
In particular, some explicit geometries and physical applications have been investigated \cite{Delduc:1993pm, Bonneau:1993wz, Howe:1991ic, Howe:1996kj, Peeters:1999ks, Ivanov:2000fg, Ivanov:2000ai, Papadopoulos:2002gy, Becker:2005nb, Kimura:2006af, Kimura:2007xa, Fedoruk:2013eua, Fedoruk:2014jba}.

It is also known that the HKT is obtained by conformal transformation of hyper-K\"{a}hler geometry \cite{Gibbons:1979zt, Callan:1991dj}. 
When the conformal factor (or the warp factor) $\e^{2 \phi}$ is a harmonic function, the HKT structure becomes strong automatically and the three-form flux can be written as $H= * \d \phi$. 
The HKTs obtained by the conformal transformation have been studied in recent years from both mathematical and physical aspects 
\cite{Gibbons:1997iy, Lambert:1997gs, Papadopoulos:1998yx, Lambert:1999ix, Grantcharov:1999kv, Papadopoulos:2000iv, Papadopoulos:2000hb, Grantcharov:2002fk, Ivanov:2010fv, Carlevaro:2013vla, Hinoue:2014csa}.

Associated with these works, we perform conformal transformation to the Taub-NUT space, a non-compact hyper-K\"{a}hler four-fold.
The Taub-NUT space represents a Kaluza-Klein five-brane (or a KK5-brane for short).
Once a conformally transformed Taub-NUT space generates an HKT whose spin connection is self-dual,
one finds that a non-trivial torsion is also encoded with the geometry.
This torsion generates a constituent of an H-monopole which is a smeared NS5-brane.
It is interesting to investigate a brane interpretation of the HKT in more detail.
If we further compactify one direction of the HKT, we obtain a reduced geometry with a fibred two-torus.
This geometry is characterized by two two-dimensional harmonic functions. 
Associated with defect branes \cite{Bergshoeff:2011se}\footnote{Recently a new interesting feature of defect branes was studied \cite{Okada:2014wma}.}, 
we refer to this geometry as the defect HKT. 
Though this has a non-trivial monodromy structure, it is unclear whether the defect HKT forms a composite state of defect five-branes or a coexistence of two defect objects.
In the present paper, we discuss this problem as a main theme. 
Furthermore, using the $O(2, 2; {\mathbb Z}) = SL(2,{\mathbb Z}) \times SL(2, {\mathbb Z})$ monodromy structure, we find a new solution that represents a defect $(p, q)$ five-brane on the ALG space \cite{Kimura:2014wga}.
The ALG space is a generalization of the ALF space discussed by \cite{Cherkis:2000cj, Cherkis:2001gm}.

The structure of this paper is as follows.
In section \ref{sect:HKT},
we briefly mention that an HKT has two harmonic functions.
One is the function which describes the structure of the Taub-NUT space.
The other is the function describing an H-monopole, a smeared NS5-brane. 
Reducing one transverse dimension in order to make two abelian isometries,
we obtain the HKT of codimension two. 
In this paper, we call them the defect HKT. 
In section \ref{sect:monodromies},
we analyze the monodromy structures of the two configurations,
where the monodromy originates from the two-torus.
We find the relations of the constituents of the defect systems.
In section \ref{sect:conjugates},
we investigate the conjugate configurations of the defect HKT and the T-dualized defect HKT.
We find new solutions which represent defect $(p,q)$ five-branes on the ALG space. 
Section \ref{sect:summary} is devoted to summary and discussions.
In appendix \ref{app:convention},
the conventions for the T-duality transformations are exhibited.
In appendix \ref{app:five-branes},
we briefly summarize the properties of various defect five-branes.

\section{A hyper-K\"{a}hler geometry with torsion}
\label{sect:HKT}

In this section, we briefly argue hyper-K\"{a}hler geometry with torsion (HKT).
We do not discuss mathematical details of HKT, and only remark that HKT is a supergravity solution. 
More detailed discussions or reviews of HKT are demonstrated in
\cite{Spindel:1988sr, Bonneau:1993wz, Howe:1996kj}. 

We begin with the background geometry of the single KK5-brane in ten-dimensional spacetime. This is given by
\bsubeq \label{KK5-system}
\begin{gather}
\d s^2_{\text{KK5}}
\ = \ 
\d s_{012345}^2
+ H_{\alpha} \big[ (\d x^6)^2 + (\d x^7)^2 + (\d x^8)^2 \big] 
+ \frac{1}{H_{\alpha}} \big[ \d y^9 - \vec{V}_{\alpha} \cdot \, \d \vec{x} \big]^2
\, , \\
B_{i9} \ = \ 0
\, , \ls
\e^{2 \phi} \ = \ 1
\, , \ls 
i \ = \ 6,7,8
\, .
\end{gather}
\esubeq
Here the vector $\vec{x}$ lives in the 678-directions.
The six-dimensional spacetime along the 012345-directions is flat, whereas the 6789-directions are described as the Taub-NUT space governed by the harmonic function $H_{\alpha}$ and the KK-vector $\vec{V}_{\alpha}$, 
whose explicit forms are given by
\bsubeq \label{HV-alpha}
\begin{gather}
H_{\alpha} \ = \ 
1 + \frac{\alpha_0}{\sqrt{2} \, |\vec{x}|}
\, , \ls
\vec{V}_{\alpha} \cdot \d \vec{x}
\ = \ 
\frac{\alpha_0}{\sqrt{2}} 
\frac{- x^6 \d x^8 + x^8 \d x^6}{|\vec{x}| (|\vec{x}| + x^7)}
\, , \\
\nabla_i H_{\alpha} \ = \ (\nabla \times \vec{V}_{\alpha})_i
\, ,
\end{gather}
\esubeq
where $\nabla_i$ is the derivative with respect to $x^i$. 
We note that $\alpha_0$ is a constant parameter.
We also comment that the 9-th direction is compactified on a circle of radius $R_9$.
This circle is called the Taub-NUT circle.
The Taub-NUT space is a hyper-K\"{a}hler geometry.
The spin connection of this four-dimensional space is self-dual.
In order to satisfy the equations of motion of ten-dimensional supergravity theories,
the B-field and the dilaton are trivial.

Now we modify the background geometry (\ref{KK5-system}) in such a way as
\bsubeq \label{HKT}
\begin{gather}
\d s^2_{\text{HKT}} 
\ \equiv \ 
\d s_{012345}^2
+ H_{\alpha} H_{\beta} \big[ (\d x^6)^2 + (\d x^7)^2 + (\d x^8)^2 \big] 
+ \frac{H_{\beta}}{H_{\alpha}} \big[ \d y^9 - \vec{V}_{\alpha} \cdot \, \d \vec{x} \big]^2
\, , \\
B_{i9} \ = \ V_{\beta,i}
\, , \ls
\e^{2 \phi} \ = \ H_{\beta}
\, .
\end{gather}
\esubeq
Here we introduced a set of functions $H_{\beta}$ and $\vec{V}_{\beta}$ whose structures are same as in (\ref{HV-alpha}), 
\bsubeq \label{HV-beta}
\begin{gather}
H_{\beta} \ = \ 
1 + \frac{\beta_0}{\sqrt{2} \, |\vec{x}|}
\, , \ls
\vec{V}_{\beta} \cdot \d \vec{x}
\ = \ 
\frac{\beta_0}{\sqrt{2}} 
\frac{- x^6 \d x^8 + x^8 \d x^6}{|\vec{x}| (|\vec{x}| + x^7)}
\, , \\
\nabla_i H_{\beta} \ = \ (\nabla \times \vec{V}_{\beta})_i
\, ,
\end{gather}
\esubeq
where $\beta_0$ is a constant parameter.
We note that the modification of the 6789-directions corresponds to the conformal transformation of the Taub-NUT metric in terms of the warp factor $\e^{2\phi} = H_{\beta}$. 

The configuration (\ref{HKT}) is also a solution to the equations of motion of supergravity theories.
We notice that the B-field and the dilaton depend on the functions $\vec{V}_{\beta}$ and $H_{\beta}$, respectively.
Due to the non-trivial B-field and the dilaton,
the space in the 6789-directions is no longer hyper-K\"{a}hler.
However, 
the new spin connection $\omega^+$ on the four-directions,
which is defined by the sum of the Levi-Civita connection $\omega$ and the three-form flux $H$, is still self-dual. 
Since the self-dual spin connection $\omega^+$ appears in the supersymmetry variations of gravitinos, the HKT solution preserves a part of supersymmetry.

We have comments on the modified geometry (\ref{HKT}).
If we set the parameter $\beta_0$ to zero,
the functions $H_{\beta}$ and $\vec{V}_{\beta}$ become trivial.
Then the modified space (\ref{HKT}) is reduced to the original one (\ref{KK5-system}).
On the other hand, if we set $\alpha_0 = 0$,
the modified geometry (\ref{HKT}) becomes the background geometry of the H-monopole, the NS5-brane smeared along the 9-th direction \cite{Gregory:1997te}.
Then we expect that the geometry (\ref{HKT}) would represent a configuration that the H-monopole governed by $H_{\beta}$ sits on the Taub-NUT space controlled by $H_{\alpha}$. 
As we will discuss later, this is indeed conceivable.
Then we think of the following question:
{\sl Is this system a bound state of the H-monopole and the KK5-brane?}
In order to find an answer of this question,
we investigate the configuration (\ref{HKT}) by virtue of the monodromy structures as discussed in \cite{Kimura:2014wga}.

As mentioned above, if one of the parameters $\alpha_0$ and $\beta_0$ vanishes, the configuration is reduced to the single defect five-brane exhibited in appendix \ref{app:five-branes}.
Then, from now on, we focus only on the configurations where both $\alpha_0$ and $\beta_0$ are non-zero.

\subsection{Defect HKTs}

In this subsection, we compactify the 8-th direction of the HKT (\ref{HKT}).
Then the two-dimensional space in the 89-directions becomes a two-torus $T^2_{89}$.
In the small radius limit of the 8-th direction, the functions
$H_{\alpha}$ and $H_{\beta}$ are deformed by the smearing procedure
\cite{deBoer:2010ud}. Then we obtain the following form,
\bsubeq \label{HKT-defect}
\begin{gather}
\d s^2_{\text{dHKT}}
\ = \ 
\d s_{012345}^2
+ H_{\alpha} H_{\beta} \big[ (\d \varrho)^2 + \varrho^2 (\d \vartheta)^2 \big] 
+ H_{\alpha} H_{\beta} (\d x^8)^2 
+ \frac{H_{\beta}}{H_{\alpha}} \big[ \d y^9 - V_{\alpha} \, \d x^8 \big]^2
\, , \\
B_{89} \ = \ V_{\beta}
\, , \ls
\e^{2 \phi} \ = \ H_{\beta}
\, .
\end{gather}
\esubeq
The geometry has the following properties.
The six-dimensional spacetime in the 012345-directions is flat.
The 67-plane is a non-compact two-dimensional space whose coordinates are reparameterized as
$x^6 = \varrho \, \cos \vartheta$ and
$x^7 = \varrho \, \sin \vartheta$, 
where $0 \leq\varrho$ and $0 \leq \vartheta \leq 2 \pi$.
The various functions now depend on the coordinates $\varrho$ and $\vartheta$ \cite{deBoer:2010ud},
\bsubeq
\begin{alignat}{3}
H_{\alpha} \ &= \ 
h + \alpha \log \frac{\mu}{\varrho}
\, , &\ls
V_{\alpha} \ &= \ 
\alpha \vartheta
\, , &\ls
\alpha \ &\equiv \ 
\frac{\alpha_0}{2 \pi R_8}
\, , \\
H_{\beta} \ &= \ 
h + \beta \log \frac{\mu}{\varrho}
\, , &\ls
V_{\beta} \ &= \ 
\beta \vartheta
\, , &\ls
\beta \ &\equiv \ 
\frac{\beta_0}{2 \pi R_8}
\, .
\end{alignat}
\esubeq
Here $R_8$ is the radius of the 8-th direction.
The functions $H_{\alpha}$ and $H_{\beta}$ are the harmonic functions on the two-dimensional 67-plane.
Thus they are logarithmic functions.
They carry the renormalization scale $\mu$ and the bare parameter $h$ which diverges in the asymptotic limit.
Due to their properties, we can only explore the system (\ref{HKT-defect}) within the well-defined region where the functions $H_{\alpha}$ and $H_{\beta}$ are positive and finite.
In the same analogy of \cite{Bergshoeff:2011se},
we refer to the geometry (\ref{HKT-defect}) as the {\sl defect} HKT.

Since the system (\ref{HKT-defect}) has two abelian isometries along the two-torus $T^2_{89}$, 
we can perform T-duality. 
If we take the T-duality transformation along the 9-th direction,
the two sets of functions $(H_{\alpha}, -V_{\alpha})$ and $(H_{\beta}, V_{\beta})$ in (\ref{HKT-defect}) are exchanged \cite{Opfermann:1997eb}.
On the other hand, if we take the T-duality transformation along the 8-th direction in (\ref{HKT-defect}),
we obtain a new configuration,
\bsubeq \label{dKK5-522-2}
\begin{gather}
\d s^2_{\text{TdHKT}} \ = \ 
\d s_{012345}^2
+ H_{\alpha} H_{\beta} \big[ (\d \varrho)^2 + \varrho^2 (\d \vartheta)^2 \big]
+ \frac{H_{\alpha}}{K_{\alpha}} \Big\{
\frac{1}{H_{\beta}} \big[ \d y^8 + V_{\beta} \, \d y^9 \big]^2 + H_{\beta} (\d y^9)^2
\Big\}
\, , \\
B_{89} \ = \ 
- \frac{V_{\alpha}}{K_{\alpha}}
\, , \ls
\e^{2 \phi} \ = \ 
\frac{H_{\alpha}}{K_{\alpha}}
\, ,
\end{gather}
\esubeq
where we have defined $y^8$ as the T-dual coordinate of $x^8$ and 
introduced
\begin{align}
K_{\alpha} 
\ &\equiv \ 
(H_{\alpha})^2 + (V_{\alpha})^2
\, , \ls
K_{\beta} 
\ \equiv \
(H_{\beta})^2 + (V_{\beta})^2
\, .
\end{align}
This is similar to the background geometry of an exotic $5^2_2$-brane (\ref{522-system}).
The crucial difference is the off-diagonal term of the metric on the fibred two-torus $T^2_{89}$.

\section{Monodromy structures}
\label{sect:monodromies}

In this section, we analyze the monodromy structures of the defect HKT (\ref{HKT-defect}) and its T-dualized system (\ref{dKK5-522-2}).
The monodromy group generated by the two-torus $T^2_{89}$ is $O(2,2;{\mathbb Z}) = SL(2,{\mathbb Z}) \times SL(2,{\mathbb Z})$.
First, we argue the $O(2,2;{\mathbb Z})$ monodromy structures of them.
Second, we further discuss the systems in terms of the $SL(2,{\mathbb Z}) \times SL(2,{\mathbb Z})$ monodromies.
In appendix \ref{app:five-branes}, 
the monodromy structures of various defect five-branes are exhibited.

\subsection{$O(2,2;{\mathbb Z})$ and $SL(2,{\mathbb Z})_{\tau} \times SL(2,{\mathbb Z})_{\rho}$ monodromies}

In the same analogy as exotic five-branes \cite{deBoer:2010ud, Kikuchi:2012za, deBoer:2012ma, Andriot:2014uda, Kimura:2014wga},
the monodromy of the defect HKT originates from the fibred two-torus $T^2_{89}$.
We package the metric $G_{mn}$ and the B-field $B_{mn}$ 
on $T^2_{89}$ in the following matrix $\mathscr{M}$,
\bsubeq \label{MM-MM}
\begin{align}
\mathscr{M} (\varrho, \vartheta) \ &\equiv \ 
\left(
\begin{array}{cc}
G_{mn} - B_{mp} \, G^{pq} \, B_{qn} & B_{mp} G^{pn} \\
- G^{mp} B_{pn} & G^{mn}
\end{array}
\right)
\, , \ls
m,n,\ldots = 8, 9
\, .
\end{align}
When we go around the origin on the 67-plane along the angular coordinate $\vartheta$,
the matrix $\mathscr{M}$ is transformed as
\begin{align}
\mathscr{M}(\varrho, 2 \pi)
\ &= \ 
\Omega^{\T} \, \mathscr{M} (\varrho, 0) \, \Omega
\, .
\end{align}
\esubeq
The transformation matrix $\Omega$ takes values in $O(2,2;{\mathbb Z})$.
We refer to this as the $O(2,2;{\mathbb Z})$ monodromy matrix.
The matrix $\Omega$ is useful to investigate (non)geometric structure of the configuration \cite{Hull:2004in, Lawrence:2006ma, Hull:2006qs, Dall'Agata:2007sr}.
In general, the matrix $\Omega$ is described 
in terms of four $2 \times 2$ blocks $A$, $D$, $\Theta$ and $\beta$ in such a way as \cite{Dall'Agata:2007sr}
\begin{align}
\Omega \ &= \ 
\left(
\begin{array}{cc}
A & \beta 
\\
\Theta & D
\end{array}
\right)
\, , \label{ADTB-O22}
\end{align}
where the block-diagonal parts $A$ and $D$ govern the coordinate transformations, 
while $\Theta$ is related to the B-field gauge transformation.
The block $\beta$ dictates the T-duality structure in the configuration.
If this block is non-trivial, the corresponding space is nongeometric.
We refer to such a configuration as a T-fold \cite{Hull:2004in}.
For instance, in the case of various defect five-branes, see \cite{Kimura:2014wga}.

Since the group $O(2,2;{\mathbb Z})$ is equivalent to $SL(2,{\mathbb Z}) \times SL(2,{\mathbb Z})$ monodromy, 
it is also worth formulating objects sensitive to the $SL(2,{\mathbb Z}) \times SL(2,{\mathbb Z})$.
Instead of the matrix $\mathscr{M}$,
we introduce two complex structures $\tau$ and $\rho$.
$\tau$ is nothing but the complex structure of the two-torus $T^2_{89}$.
This is subject to the first $SL(2,{\mathbb Z})$.
On the other hand, $\rho$ is defined by the B-field $B_{mn}$ and the metric $G_{mn}$ in the following way \cite{deBoer:2012ma},
\bsubeq \label{field-config}
\begin{gather}
\rho \ \equiv \ 
B_{89} + \I \sqrt{\det G_{mn}}
\, .
\end{gather}
The complex structure $\rho$ 
controls the scale of the two-torus $T^2_{89}$.
This is subject to the second $SL(2,{\mathbb Z})$.
By virtue of the two complex structures $\tau$ and $\rho$, 
we can represents $G_{mn}$, $B_{89}$ and $\phi$,
\begin{gather}
G_{mn} \ = \ 
\frac{\rho_2}{\tau_2} \left(
\begin{array}{cc}
1 & \tau_1
\\
\tau_1 & |\tau|^2
\end{array}
\right)
\, , \ls
B_{89} \ = \ \rho_1
\, , \ls
\e^{2 \phi} \ = \ \rho_2
\, ,
\end{gather}
\esubeq
where we decomposed $\tau = \tau_1 + \I \tau_2$ and $\rho = \rho_1 + \I \rho_2$.
In the rest of this section, 
we analyze both the $O(2,2;{\mathbb Z})$ monodromies and the $SL(2,{\mathbb Z})_{\tau} \times SL(2,{\mathbb Z})_{\rho}$ monodromies of the defect HKT (\ref{HKT-defect}) and its T-dualized configuration (\ref{dKK5-522-2}).

\subsection{Monodromies of the defect HKT}

In this subsection, we study the monodromy structures of the defect HKT.
The explicit description of the matrix $\mathscr{M}$ (\ref{MM-MM}) for the configuration (\ref{HKT-defect}) 
is 
\bsubeq \label{MM-MM-dHKT}
\begin{align}
\mathscr{M}^{\text{dHKT}} (\varrho,\vartheta)
\ &= \ 
\frac{1}{H_{\alpha} H_{\beta}} \left(
\begin{array}{cccc}
K_{\alpha} K_{\beta} & - V_{\alpha} K_{\beta} & V_{\alpha} V_{\beta} & K_{\alpha} V_{\beta}
\\
- V_{\alpha} K_{\beta} & K_{\beta} & - V_{\beta} & - V_{\alpha} V_{\beta}
\\
V_{\alpha} V_{\beta} & - V_{\beta} & 1 & V_{\alpha}
\\
K_{\alpha} V_{\beta} & - V_{\alpha} V_{\beta} & V_{\alpha} & K_{\alpha}
\end{array}
\right)
\, . \label{moduliM-dHKT} 
\end{align}
This is rather a complicated. 
However, the $O(2,2;{\mathbb Z})$ monodromy matrix $\Omega$ is a simple form,
\begin{align}
\Omega^{\text{dHKT}}_{\alpha,\beta}
\ &= \ 
\left(
\begin{array}{cccc}
1 & 0 & 0 & 0 
\\
- 2 \pi \alpha & 1 & 0 & 0 
\\
(2 \pi)^2 {\alpha} {\beta} & - 2 \pi {\beta} & 1 & 2 \pi {\alpha}
\\
2 \pi {\beta} & 0 & 0 & 1
\end{array}
\right)
\, . \label{monodromyM-dHKT} 
\end{align}
\esubeq
We immediately find a relation
$\Omega_{\alpha,\beta}^{\text{dHKT}} = 
\Omega^{\text{KK}}_{\alpha} \Omega^{\text{NS}}_{\beta}
= \Omega^{\text{NS}}_{\beta} \Omega^{\text{KK}}_{\alpha}$,
where the explicit forms of $\Omega^{\text{NS}}_{\beta}$ and $\Omega^{\text{KK}}_{\alpha}$ are given in appendix \ref{app:five-branes}.
This seems that the defect NS5-brane in (\ref{HKT-defect}) is not influenced by the defect KK5-brane.
This would also be interpreted that the defect Taub-NUT space\footnote{For this terminology, see appendix \ref{app:five-branes}.} does not receive any back reactions from the defect NS5-brane.
Due to this fact, we expect that the H-monopole and the KK5-brane do not combine to form a single composite object in the original configuration of codimension three (\ref{HKT}).

Through the investigation of the $SL(2,{\mathbb Z})_{\tau} \times SL(2,{\mathbb Z})_{\rho}$ monodromy structure, 
we can further study the relation between the two different defect five-branes.
The two complex structures $\tau$ and $\rho$ are described as
\bsubeq \label{tr-dHKT}
\begin{align}
\tau \ &= \ 
\frac{-1}{V_{\alpha} + \I H_{\alpha}}
\ = \
\frac{\I}{h + \alpha \log (\mu/z)}
\, , \\
\rho \ &= \ 
V_{\beta} + \I H_{\beta}
\ = \ 
\I h + \I \beta \log (\mu/z) 
\, . 
\end{align}
\esubeq
Here we defined the complex coordinate $z = \varrho \, \e^{\I \vartheta}$ in the 67-plane.
When we go around the origin in the 67-plane $z \to z \, \e^{2 \pi \I}$,
these complex structures are transformed, 
\bsubeq \label{dHKT-SL2SL2-monodromy}
\begin{alignat}{2}
\tau \ &\to \ 
\tau' \ = \ 
\frac{\tau}{- 2 \pi \alpha \tau + 1}
\, , &\ls
\Omega_{\tau}^{\text{dHKT}}
\ &\equiv \ 
\left(
\begin{array}{cc}
1 & 0 \\
- 2 \pi \alpha & 1
\end{array}
\right)
\, , \\
\rho \ &\to \ 
\rho' \ = \ \rho + 2 \pi \beta
\, , &\ls
\Omega_{\rho}^{\text{dHKT}}
\ &\equiv \ 
\left(
\begin{array}{cc}
1 & + 2 \pi \beta \\
0 & 1
\end{array}
\right)
\, .
\end{alignat}
\esubeq
This denotes that the structure of the defect HKT is quite different from that of the single defect five-branes (see appendix \ref{app:five-branes}).
The monodromy structures of the defect HKT are generated by both $\tau$ and $\rho$, 
whereas the monodromies of the single defect five-branes are given only by one of them.
We notice that the monodromy matrices $\Omega_{\tau}^{\text{dHKT}}$ and $\Omega_{\rho}^{\text{dHKT}}$ are independent of each other.
We will discuss the physical interpretation of these results later.

\subsection{Monodromies of the T-dualized defect HKT}

We also analyze the monodromy structures of the T-dualized defect HKT (\ref{dKK5-522-2}).
The matrices $\mathscr{M}$ and $\Omega$ are explicitly expressed as
\bsubeq \label{MM-MM-TdHKT}
\begin{align}
\mathscr{M}^{\text{TdHKT}} (\varrho,\vartheta)
\ &= \ 
\frac{1}{H_{\alpha} H_{\beta}}
\left(
\begin{array}{cccc}
1 & V_{\beta} & V_{\alpha} V_{\beta} & - V_{\alpha}
\\
V_{\beta} & K_{\beta} & V_{\alpha} K_{\beta}  & - V_{\alpha} V_{\beta}
\\
V_{\alpha} V_{\beta} & V_{\alpha} K_{\beta}  & K_{\alpha} K_{\beta} & - K_{\alpha} V_{\beta}
\\
- V_{\alpha} & - V_{\alpha} V_{\beta} & - K_{\alpha} V_{\beta} & K_{\alpha}
\end{array}
\right)
\, , \label{moduliM-TdHKT} \\
\Omega^{\text{TdHKT}}_{\alpha,\beta} 
\ &= \ 
\left(
\begin{array}{cccc}
1 & 2 \pi {\beta} & (2 \pi)^2 {\alpha} {\beta} & - 2 \pi {\alpha}
\\
0 & 1 & 2 \pi {\alpha} & 0 
\\
0 & 0 & 1 & 0 
\\
0 & 0 & - 2 \pi {\beta} & 1
\end{array}
\right)
\, . \label{monodromyM-TdHKT}
\end{align}
\esubeq
We again find a simple relation of the $O(2,2;{\mathbb Z})$ monodromy matrices 
$\Omega_{\alpha,\beta}^{\text{TdHKT}} = \Omega^{\text{AK}}_{\beta} \Omega^{\text{E}}_{\alpha} = \Omega^{\text{E}}_{\alpha} \Omega^{\text{AK}}_{\beta}$.
This indicates that, in the configuration (\ref{dKK5-522-2}), the constituent given by $H_{\alpha}$ is again independent of the constituent governed by $H_{\beta}$.
We also find that the $\Omega_{\alpha,\beta}^{\text{TdHKT}}$ possesses the $2 \times 2$ block $\beta$ in (\ref{ADTB-O22}).
Hence the T-dualized defect HKT is a T-fold.

We also analyze the system by virtue of the $SL(2,{\mathbb Z})_{\tau} \times SL(2,{\mathbb Z})_{\rho}$ monodromy structure.
The two complex structures are described as
\bsubeq \label{tr-TdHKT}
\begin{align}
\tau \ &= \ 
V_{\beta} + \I H_{\beta}
\ = \ 
\I h + \I \beta \log (\mu/z)
\, , \\
\rho \ &= \ 
\frac{-1}{V_{\alpha} + \I H_{\alpha}}
\ = \ 
\frac{\I}{h + \alpha \log (\mu/z)}
\, . 
\end{align}
\esubeq
Compared to the formulation (\ref{tr-dHKT}),
we find that the T-duality transformation exchanges the two complex structures.
Under the shift $z \to z\, \e^{2 \pi \I}$,
their monodromy matrices are given as
\bsubeq \label{TdHKT-SL2SL2-monodromy}
\begin{alignat}{2}
\tau \ &\to \ 
\tau' \ = \  \tau + 2 \pi \beta 
\, , &\ls
\Omega_{\tau}^{\text{TdHKT}}
\ &\equiv \ 
\left(
\begin{array}{cc}
1 & + 2 \pi \beta \\
0 & 1
\end{array}
\right)
\, , \\
\rho \ &\to \ 
\rho' \ = \ 
\frac{\rho}{- 2 \pi \alpha \rho + 1}
\, , &\ls
\Omega_{\rho}^{\text{TdHKT}}
\ &\equiv \ 
\left(
\begin{array}{cc}
1 & 0 \\
- 2 \pi \alpha & 1
\end{array}
\right)
\, .
\end{alignat}
\esubeq
Again we find that the monodromy matrices $\Omega_{\tau}^{\text{TdHKT}}$ and $\Omega_{\rho}^{\text{TdHKT}}$ are equal to those of the defect KK5-brane of another type and of the exotic $5^2_2$-brane, respectively (see \cite{Kimura:2014wga} and appendix \ref{app:five-branes}).
The nongeometric structure of the T-dualized defect HKT is generated by $H_{\alpha}$, i.e., the property of the exotic $5^2_2$-brane. 

\subsection{Composites as $(p,q)$ five-branes?}

The monodromy matrix $\Omega_{\tau}^{\text{dHKT}}$ is the same as that of the single defect KK5-brane, 
whereas the matrix $\Omega_{\rho}^{\text{dHKT}}$ is equal to that of the single defect NS5-brane.  
Hence we can interpret that the two constituents from $H_{\beta}$ and $H_{\alpha}$ in (\ref{HKT-defect}) stand as the independent defect NS5-brane and defect KK5-brane, respectively.
This leads us that the original configuration (\ref{HKT}) of codimension three is {\sl not} a bound state of the H-monopole and the KK5-brane {\sl in a sense of} $(p,q)$ five-branes or $(p,q)$ seven-branes in type IIB theory \cite{Bergshoeff:2006jj}.
Thus we would be able to think of (\ref{HKT}) or (\ref{HKT-defect}) as a coexistent state.
The same situation can be seen in the T-dualized system (\ref{dKK5-522-2}), 
where the single defect KK5-brane of another type and the single exotic $5^2_2$-brane are independent constituents.

In the next section, we will further discuss this issue from the perspective of the conjugate configurations.

\section{Conjugate configurations}
\label{sect:conjugates}

In the previous section, we have studied the HKT solution and its 
monodromy structures in ten-dimensional supergravity compactified on the fibred two-torus $T^2_{89}$.
In general, ten-dimensional string theory compactified on the two-torus possesses the $O(2,2;\mathbb{Z})$ T-duality structure.
We therefore expect a family of solutions which belong to the $O(2,2;\mathbb{Z})$ multiplet. 
When a solution associated with the monodromy matrix $\Omega$ exists,
there is a conjugate solution whose monodromy matrix is given by $U^{-1} \Omega U$, where $U$ is a certain $O(2,2;\mathbb{Z})$ matrix.
This strategy is completely parallel to that for seven-branes in type IIB string theory \cite{Bergshoeff:2006jj}, and for defect five-branes \cite{Hellerman:2002ax, Kimura:2014wga}.

In this section, we investigate conjugate configurations of the defect HKT and its T-dualized system.
We introduce the following general matrices of $SL(2,{\mathbb Z})_{\tau} \times SL(2,{\mathbb Z})_{\rho}$,
\bsubeq \label{conjugate-U}
\begin{gather}
U_{\tau} \ \equiv \ 
\left(
\begin{array}{cc}
s' & r' \\
q' & p'
\end{array}
\right)
\, , \ls
U_{\rho} \ \equiv \ 
\left(
\begin{array}{cc}
s & r \\
q & p
\end{array}
\right)
\, , \ls
\begin{array}{rcl}
s'p' - r'q' \!\!&=&\!\! 1
\, , \\
sp - rq \!\!&=&\!\! 1
\, .
\end{array}
\end{gather}
Here we note that all components of the matrices are integer.
We transform the $SL(2,{\mathbb Z})_{\tau} \times SL(2,{\mathbb Z})_{\rho}$ monodromy matrices (\ref{dHKT-SL2SL2-monodromy}) and  (\ref{TdHKT-SL2SL2-monodromy}), 
\begin{align}
\Omega_{\tau,\rho} \ \to \ 
\wt{\Omega}_{\tau,\rho} \ = \ 
U_{\tau,\rho}^{-1} \Omega_{\tau,\rho}^{\vphantom{-1}} U_{\tau,\rho}^{\vphantom{-1}}
\, .
\end{align}
\esubeq
Then we construct a family of solutions associated with the monodromy matrices $\wt{\Omega}_{\tau,\rho}$.
We have already known that this technique is also useful to study bound states of defect five-branes \cite{deBoer:2010ud, Kikuchi:2012za, deBoer:2012ma, Kimura:2014wga, Okada:2014wma}.

\subsection{Conjugate of defect HKT}

First, we apply the conjugate procedure (\ref{conjugate-U}) to the $SL(2,{\mathbb Z})_{\tau} \times SL(2,{\mathbb Z})_{\rho}$ monodromy matrices for the defect HKT (\ref{dHKT-SL2SL2-monodromy}).
The transformed matrices are
\bsubeq \label{conjugate-dHKT}
\begin{align}
\wt{\Omega}_{\tau}^{\text{dHKT}}
\ &= \ 
\left(
\begin{array}{cc}
1 + 2 \pi \alpha r's' & 2 \pi \alpha r'{}^2 
\\
- 2 \pi \alpha s'{}^2 & 1 - 2 \pi \alpha r's'
\end{array}
\right)
\, , \\
\wt{\Omega}_{\rho}^{\text{dHKT}}
\ &= \ 
\left(
\begin{array}{cc}
1 + 2 \pi \beta pq & 2 \pi \beta p^2 
\\
- 2 \pi \beta q^2 & 1 - 2 \pi \beta pq
\end{array}
\right)
\, .
\end{align}
\esubeq
Since $\alpha$ and $\beta$ are non-vanishing,
the conjugate of the defect HKT is different from any conjugates of single defect five-brane in appendix \ref{app:five-branes}.
Thus we again understand that the defect HKT system is a coexistent state of defect NS5-branes and defect KK5-branes.
This is completely different from defect $(p,q)$ five-branes. 

It is worth exploring the conjugate configuration of the defect HKT.
Applying the conjugate transformations (\ref{conjugate-U}) to the two complex structures (\ref{tr-dHKT}),
we obtain the conjugate complex structures $\wt{\tau}$ and $\wt{\rho}$ in the following way \cite{Kimura:2014wga},
\bsubeq \label{conjugate-tr-dHKT}
\begin{alignat}{2}
U_{\tau}^{-1} \ &= \ 
\left(
\begin{array}{cc}
p' & -r' \\
-q' & s'
\end{array}
\right)
\, , &\ls
\tau \ \to \ 
\wt{\tau} \ &\equiv \
\frac{p' \tau - r'}{-q' \tau + s'}
\, , \label{conjugate-tau-dHKT} \\
U_{\rho}^{-1} \ &= \ 
\left(
\begin{array}{cc}
p & -r \\
- q & s
\end{array}
\right)
\, , &\ls
\rho \ \to \ 
\wt{\rho} \ &\equiv \
\frac{p \rho - r}{-q \rho + s}
\, . \label{conjugate-rho-dHKT}
\end{alignat}
\esubeq
Going around the origin in the 67-plane $z \to z \, \e^{2 \pi \I}$,
the original complex structures $\tau$ and $\rho$ are transformed as (\ref{dHKT-SL2SL2-monodromy}).
Then we find that the complex structures $\wt{\tau}$ and $\wt{\rho}$ are also transformed as 
\bsubeq 
\begin{align}
\wt{\tau} \ \to \ 
\wt{\tau}' \ &= \ 
\frac{(1 + 2 \pi \alpha r' s') \wt{\tau} + 2 \pi \alpha r'{}^2}{-2 \pi \alpha s'{}^2 \wt{\tau} + (1 - 2 \pi \alpha r' s')}
\, , \\
\wt{\rho} \ \to \ 
\wt{\rho}' \ &= \ 
\frac{(1 + 2 \pi \beta pq) \wt{\rho} + 2 \pi \beta p^2}{- 2 \pi \beta q^2 \wt{\rho} + (1 - 2 \pi \beta pq)}
\, .
\end{align}
\esubeq
We can immediately confirm that the conjugate complex structures reproduce the conjugate monodromy matrices (\ref{conjugate-dHKT}).
The explicit expression of the conjugate complex structures is
\bsubeq \label{tau-rho-conjugate-dHKT}
\begin{align}
\wt{\tau} \ &= \ 
- \frac{r' (V_{\alpha} + \I H_{\alpha}) + p'}{s' (V_{\alpha} + \I H_{\alpha}) + q'}
\ = \ 
\frac{-[p'q' + (p's' + r'q') V_{\alpha} + r' s' K_{\alpha}] + \I H_{\alpha}}{q'{}^2 + 2 q's' V_{\alpha} + s'{}^2 K_{\alpha}}
\, ,  \\
\wt{\rho} \ &= \ 
\frac{p (V_{\beta} + \I H_{\beta}) - r}{- q (V_{\beta} + \I H_{\beta}) -s}
\ = \ 
\frac{[-rs + (ps + rq) V_{\beta} - pq K_{\beta}] + \I H_{\beta}}{s^2 - 2 qs V_{\beta} + q^2 K_{\beta}}
\, . 
\end{align}
\esubeq
Substituting (\ref{tau-rho-conjugate-dHKT}) into the formulation (\ref{field-config}), we obtain the explicit form of $\wt{G}_{mn}$, $\wt{B}_{89}$ and $\wt{\phi}$,
\bsubeq \label{conjugate-BGphi-dHKT}
\begin{align}
\wt{G}_{88} 
\ &= \ 
\frac{H_{\beta}}{H_{\alpha}}
\frac{q'{}^2 + 2 q's' V_{\alpha} + s'{}^2 K_{\alpha}}{s^2 - 2 qs V_{\beta} + q^2 K_{\beta}}
\, , \\
\wt{G}_{89} \ = \ \wt{G}_{98} 
\ &= \ 
- \frac{H_{\beta}}{H_{\alpha}}
\frac{p'q' + (p's' + r'q') V_{\alpha} + r' s' K_{\alpha}}{s^2 - 2 qs V_{\beta} + q^2 K_{\beta}}
\, , \\
\wt{G}_{99} 
\ &= \ 
\frac{H_{\beta}}{H_{\alpha}}
\frac{p'{}^2 + 2 p'r' V_{\alpha} + r'{}^2 K_{\alpha}}{s^2 - 2 qs V_{\beta} + q^2 K_{\beta}}
\, , \\
\wt{B}_{89} \ &= \ 
- \frac{rs - (ps + rq) V_{\beta} + pq K_{\beta}}{s^2 - 2 qs V_{\beta} + q^2 K_{\beta}}
\, , \\
\e^{2 \wt{\phi}}
\ &= \ 
\frac{H_{\beta}}{s^2 - 2 qs V_{\beta} + q^2 K_{\beta}}
\, .
\end{align}
\esubeq
This is the generic configuration.
However, this is rather a lengthy expression because redundant parameters are included.
Now we look for genuinely a new configuration under the non-vanishing parameters $(r',s')$ and $(p,q)$.
If not, the conjugate system is reduced to the original one (\ref{HKT-defect}) or (\ref{dKK5-522-2}).
For convenience, we introduce a new variable $\wh{\lambda} = - 1/\wh{\tau}$.
By using the constraints $s'p' - q'r' = 1$ and $sp-qr=1$ in (\ref{conjugate-U}),
we can remove redundant parameters $p'$ and $s$,
\bsubeq \label{taurho2-dHKT}
\begin{align}
\wh{\lambda} 
\ &= \ 
\frac{s'}{r'} - \frac{1}{r'{}^2 (V_{\alpha} + \I H_{\alpha})}
\, , \\
\wh{\rho} \ &= \ 
- \frac{p}{q} - \frac{1}{q^2 (V_{\beta} + \I H_{\beta})}
\, .
\end{align}
\esubeq
Under the shift $z \to z \, \e^{2 \pi \I}$, they can again realize the conjugate monodromy matrices (\ref{conjugate-dHKT}).
Then, substituting (\ref{taurho2-dHKT}) in the formulation (\ref{field-config}),
we obtain the simplified field configuration,
\bsubeq \label{ConjugateConfig-dHKT}
\begin{gather}
\d s^2 \ = \ 
\d s_{012345}^2 
+ H_{\alpha} H_{\beta} \big[ (\d \varrho)^2 + \varrho^2 (\d \vartheta)^2 \big]
+ \wh{\lambda}_2 \wh{\rho}_2 \, (\d x^8)^2
+ \frac{\wh{\rho}_2}{\wh{\lambda}_2} \big[ \d y^9 - \wh{\lambda}_1 \, \d x^8 \big]^2 
\, , \\
\wh{B}_{89} \ = \ \wh{\rho}_1 
\, , \ls
\e^{2 \wh{\phi}} \ = \ \wh{\rho}_2 
\, .
\end{gather}
Here we utilized $\wh{\lambda} = -1/\wh{\tau} \equiv \wh{\lambda}_1 + \I \wh{\lambda}_2$ which is much more useful to express the configuration than $\wh{\tau}$ itself.
The components of $\wh{\lambda}$ and $\wh{\rho}$ are
\begin{alignat}{2}
\wh{\lambda}_1 \ &= \ \frac{s'}{r'} - \frac{V_{\alpha}}{r'{}^2 K_{\alpha}}
\, , &\ls 
\wh{\lambda}_2 \ &= \ \frac{H_{\alpha}}{r'{}^2 K_{\alpha}}
\, , \\
\wh{\rho}_1 \ &= \ - \frac{p}{q} - \frac{V_{\beta}}{q^2 K_{\beta}}
\, , &\ls 
\wh{\rho}_2 \ &= \ \frac{H_{\beta}}{q^2 K_{\beta}}
\, .
\end{alignat}
\esubeq
We can check that the configuration (\ref{ConjugateConfig-dHKT}) is a solution of supergravity theories.

We discuss a physical interpretation of this solution as follows.
Before the conjugate, we recognize that the defect HKT corresponds to the single defect NS5-brane on the defect Taub-NUT space.
The defect NS5-brane and the defect Taub-NUT space are not affected by each other.
Then, after the conjugate transformation,
the defect NS5-brane is transformed to a composite state, i.e.,
a composite of $p$ defect NS5-branes and $q$ exotic $5^2_2$-branes, called the defect $(p,q)$ five-brane \cite{Kimura:2014wga}.
On the other hand, the defect Taub-NUT sector, equivalent to the single defect KK5-brane, is also transformed to a composite of $-s'$ defect KK5-branes and $r'$ defect KK5-branes of another type.
This can be the ALG space \cite{Cherkis:2000cj, Cherkis:2001gm, Hellerman:2002ax, Kimura:2014wga}. 
Then we conclude that the conjugate configuration (\ref{ConjugateConfig-dHKT}) is a defect $(p,q)$ five-brane on the ALG space.

\subsection{Conjugate of T-dualized defect HKT}

In this subsection, we study the conjugate of the T-dualized defect HKT (\ref{dKK5-522-2}).
The $SL(2,{\mathbb Z})_{\tau} \times SL(2,{\mathbb Z})_{\rho}$ monodromy matrices (\ref{TdHKT-SL2SL2-monodromy}) are transformed in terms of the conjugate matrices (\ref{conjugate-U}),
\bsubeq \label{conjugate-TdHKT}
\begin{align}
\wt{\Omega}_{\tau}^{\text{TdHKT}}
\ &= \ 
\left(
\begin{array}{cc}
1 + 2 \pi \beta p'q' & 2 \pi \beta p'{}^2 
\\
- 2 \pi \beta q'{}^2 & 1 - 2 \pi \beta p'q'
\end{array}
\right)
\, , \\
\wt{\Omega}_{\rho}^{\text{TdHKT}}
\ &= \ 
\left(
\begin{array}{cc}
1 + 2 \pi \alpha rs & 2 \pi \alpha r^2 
\\
- 2 \pi \alpha s^2 & 1 - 2 \pi \alpha rs
\end{array}
\right)
\, .
\end{align}
\esubeq
Since both $\alpha$ and $\beta$ are non-vanishing,
the above matrices are never reduced to those of (the conjugate of) the defect five-branes in appendix \ref{app:five-branes}.
Applying the conjugate transformation to the two complex structures, 
we obtain the conjugates $\wt{\tau}$ and $\wt{\rho}$,
\bsubeq \label{conjugate-tr-TdHKT}
\begin{alignat}{2}
U_{\tau}^{-1} \ &= \ 
\left(
\begin{array}{cc}
p' & -r' \\
- q' & s'
\end{array}
\right)
\, , &\ls
\tau \ \to \ 
\wt{\tau} \ &\equiv \
\frac{p' \tau - r'}{-q' \tau + s'}
\, , \label{conjugate-tau-TdHKT} \\
U_{\rho}^{-1} \ &= \ 
\left(
\begin{array}{cc}
p & -r \\
-q & s
\end{array}
\right)
\, , &\ls
\rho \ \to \ 
\wt{\rho} \ &\equiv \
\frac{p \rho - r}{- q \rho + s}
\, . \label{conjugate-rho-TdHKT}
\end{alignat}
\esubeq
Since the original complex structures $\tau$ and $\rho$ are transformed as
(\ref{TdHKT-SL2SL2-monodromy}) under the shift $z \to z \, \e^{2 \pi \I}$,
the conjugate complex structures are transformed in such a way as
\bsubeq
\begin{align}
\wt{\tau} \ \to \ 
\wt{\tau}' \ &= \ 
\frac{(1 + 2 \pi \beta p' q') \wt{\tau} + 2 \pi \beta p'{}^2}{- 2 \pi \beta q'{}^2 \wt{\tau} + (1 - 2 \pi \beta p'q')}
\, , \\
\wt{\rho} \ \to \ 
\wt{\rho}' \ &= \ 
\frac{(1 + 2 \pi \alpha rs) \wt{\rho} + 2 \pi \alpha r^2}{- 2 \pi \alpha s^2 \wt{\rho} + (1 - 2 \pi \alpha rs)}
\, .
\end{align}
\esubeq
They truly reproduce the conjugate monodromy matrices (\ref{conjugate-TdHKT}).
Here we write the explicit form of the conjugate complex structures,
\bsubeq \label{tau-rho-conjugate-TdHKT}
\begin{align}
\wt{\tau} \ &= \ 
\frac{p' (V_{\beta} + \I H_{\beta}) - r'}{- q' (V_{\beta} + \I H_{\beta}) + s'}
\ = \
\frac{[- r's' + (p's' + r'q') V_{\beta} - p'q' K_{\beta}] + \I H_{\beta}}{s'{}^2 - 2 q' s' V_{\beta} + q'{}^2 K_{\beta}}
\, ,  \\
\wt{\rho} \ &= \ 
- \frac{r (V_{\alpha} + \I H_{\alpha}) + p}{s (V_{\alpha} + \I H_{\alpha}) + q}
\ = \ 
\frac{- [pq + (ps + rq) V_{\alpha} + rs K_{\alpha}] + \I H_{\alpha}}{q^2 + 2 qs V_{\alpha} + s^2 K_{\alpha}}
\, .
\end{align}
\esubeq
Hence, plugging (\ref{tau-rho-conjugate-TdHKT}) into (\ref{field-config}),
we obtain the generic form of $\wt{G}_{mn}$, $\wt{B}_{89}$ and $\wt{\phi}$,
\bsubeq \label{conjugate-BGphi-TdHKT}
\begin{align}
\wt{G}_{88} 
\ &= \ 
\frac{H_{\alpha}}{H_{\beta}}
\frac{s'{}^2 - 2 q' s' V_{\beta} + q'{}^2 K_{\beta}}{q^2 + 2 qs V_{\alpha} + s^2 K_{\alpha}}
\, , \\
\wt{G}_{89} \ = \ \wt{G}_{98} 
\ &= \ 
- \frac{H_{\alpha}}{H_{\beta}}
\frac{r's' - (p's' + r'q') V_{\beta} + p'q' K_{\beta}}{q^2 + 2 qs V_{\alpha} + s^2 K_{\alpha}}
\, , \\
\wt{G}_{99} 
\ &= \ 
\frac{H_{\alpha}}{H_{\beta}}
\frac{r'{}^2 - 2 p' r' V_{\beta} + p'{}^2 K_{\beta}}{q^2 + 2 qs V_{\alpha} + s^2 K_{\alpha}}
\, , \\
\wt{B}_{89} \ &= \ 
- \frac{pq + (ps + rq) V_{\alpha} + rs K_{\alpha}}{q^2 + 2 qs V_{\alpha} + s^2 K_{\alpha}}
\, , \\
\e^{2 \wt{\phi}} \ &= \ 
\frac{H_{\alpha}}{q^2 + 2 qs V_{\alpha} + s^2 K_{\alpha}}
\, .
\end{align}
\esubeq
This is again a redundant expression.
Here we focus on the situation that the parameters $(p',q')$ and $(r,s)$ are non-vanishing, otherwise the configuration is reduced to the original one (\ref{HKT-defect}) or (\ref{dKK5-522-2}).
It is convenient to introduce $\wh{\omega} = - 1/\wh{\rho}$.
By virtue of the constraints in (\ref{conjugate-U}), 
we can remove redundant parameters $s'$ and $p$ without loss of generality.
The simplified complex structures,
\bsubeq \label{taurho2-TdHKT}
\begin{align}
\wh{\tau} 
\ &= \ 
- \frac{p'}{q'} - \frac{1}{q'{}^2 (V_{\beta} + \I H_{\beta})}
\, , \\
\wh{\omega} 
\ &= \ 
\frac{s}{r} - \frac{1}{r^2 (V_{\alpha} + \I H_{\alpha})}
\, ,
\end{align}
\esubeq
again reproduce the conjugate monodromy matrices (\ref{conjugate-TdHKT}).
Substituting (\ref{taurho2-TdHKT}) into (\ref{field-config}),
we obtain the reduced form of the conjugate configuration,
\bsubeq \label{ConjugateConfig-TdHKT}
\begin{gather}
\d s^2 \ = \ 
\d s_{012345}^2 
+ H_{\alpha} H_{\beta} \big[ (\d \varrho)^2 + \varrho^2 (\d \vartheta)^2 \big]
+ \frac{\wh{\omega}_2}{\wh{\tau}_2 |\wh{\omega}|^2} \big[ \d y^8 + \wh{\tau}_1 \, \d y^9 \big]^2
+ \frac{\wh{\tau}_2 \wh{\omega}_2}{|\wh{\omega}|^2} (\d y^9)^2
\, , \\
\wh{B}_{89} \ = \ - \frac{\wh{\omega}_1}{|\wh{\omega}|^2}
\, , \ls
\e^{2 \wh{\phi}} \ = \ \frac{\wh{\omega}_2}{|\wh{\omega}|^2}
\, .
\end{gather}
Here we used $\wh{\omega} = -1/\wh{\rho} \equiv \wh{\omega}_1 + \I \wh{\omega}_2$.
The explicit expression of the components is
\begin{alignat}{2}
\wh{\tau}_1 \ &= \ - \frac{p'}{q'} - \frac{V_{\beta}}{q'{}^2 K_{\beta}}
\, , &\ls 
\wh{\tau}_2 \ &= \ \frac{H_{\beta}}{q'{}^2 K_{\beta}}
\, , \\
\wh{\omega}_1 \ &= \ \frac{s}{r} - \frac{V_{\alpha}}{r^2 K_{\alpha}}
\, , &\ls 
\wh{\omega}_2 \ &= \ \frac{H_{\alpha}}{r^2 K_{\alpha}}
\, .
\end{alignat}
\esubeq
We can confirm that (\ref{ConjugateConfig-TdHKT}) is also a solution of supergravity theories.
This is also obtained from the previous solution (\ref{ConjugateConfig-dHKT}) via the T-duality transformation along the 8-th direction and relabeling $(s',r') = (s,r)$ and $(p,q) = (p',q')$.
The physical interpretation of this configuration is a defect $(r,-s)$ five-brane dictated by $H_{\alpha}$ on the ALG space governed by $H_{\beta}$.
This ALG space also can be regarded as a composite of $q'$ defect KK5-branes and $p'$ defect KK5-branes of another type.

\section{Summary and discussions}
\label{sect:summary}

In this paper we found new solutions associated with hyper-K\"{a}hler
geometry with torsion (HKT) and studied their monodromy structures.
The solutions represent coexistent two parallel five-branes.

We started from the single KK5-brane solution of codimension three.
This is purely a geometrical solution, namely, only the metric is non-trivial, whereas the B-field and the dilaton vanish.
The transverse direction of the KK5-brane is given by the Taub-NUT space
whose center is specified by the harmonic function $H_{\alpha}$
and the geometry is hyper-K\"{a}hler.
We then considered a new field configuration where the non-trivial B-field is turned on along the transverse directions to the single KK5-brane.
This geometry is called HKT, a conformally transformed Taub-NUT space. 
The HKT-connection which consists of the Levi-Civita connection and the three-form flux $H$ preserves self-duality.
We found that the field configuration is a solution to ten-dimensional supergravity theories, 
provided that a non-zero torsion in the background geometry is introduced. 
The torsion is governed by another harmonic function $H_{\beta}$ and 
the geometry is deformed from the Taub-NUT space.
The B-field does not belong to pure gauge 
and therefore the resultant HKT solution \eqref{HKT} is not gauge equivalent to the KK5-brane geometry.
The new solution \eqref{HKT} incorporates isometries along not only the flat six-dimensional directions longitudinal to the KK5-brane worldvolume but also one of the transverse directions.
Therefore the new solution represents a stuck of five-branes 
and they are magnetic sources of the B-field.
The isometry along the transverse direction is a reminiscent of the Taub-NUT space.

In order to understand physical properties of the new solution,
studying monodromy structures of the solutions is useful.
To this end, we compactified the geometry along the transverse 8-th direction
which is different from the isometry direction of the Taub-NUT space.
Now the five-branes become defect branes of codimension two and 
there are two isometries along the transverse directions.
The geometry is governed by the two logarithmic harmonic functions
$H_{\alpha}$ and $H_{\beta}$ which are no longer single-valued.
We call this geometry the defect HKT. 
This geometry is characterized by the $O(2,2;\mathbb{Z}) = SL(2,\mathbb{Z}) \times SL(2,\mathbb{Z})$ monodromy originated from the fibred two-torus $T^2_{89}$.

First, we calculated the monodromy matrix $\Omega$ associated with the new solution. 
We found that the monodromy matrix $\Omega_{\alpha,\beta}^{\text{dHKT}}$ is
decomposed into a product of those of the defect KK5-brane $\Omega_{\alpha}^{\text{KK}}$ and the defect NS5-brane $\Omega_{\beta}^{\text{NS}}$. 
We found that these two monodromy matrices are commutative with each other.
Next, we performed the T-duality transformation of the defect HKT solution along the 8-th direction.
We found that the T-dualized monodromy matrix is given by a product of those for another defect KK5-brane and an exotic $5^2_2$-brane. 
We then found that the solution \eqref{HKT-defect} is not connected to any single defect five-brane configurations. 
Therefore the monodromy structures suggest that the defect HKT solution,
and its higher dimensional origin \eqref{HKT}, represent a coexistent (defect) five-branes different from a composite state of two (defect) five-branes.

Since ten-dimensional supergravity theories compactified on the two-torus $T^2_{89}$
have the $O(2,2;{\mathbb Z})$ T-duality structure, we expected that there is a family of five-brane solutions.  
This situation is similar to defect $(p,q)$ $n$-branes in type IIB string
theory\footnote{For $n=1,5,7$, they are bound states of $p$ D-strings
and $q$ F-strings, $p$ D5-branes and $q$ NS5-branes, $p$ D7-branes and
$q$ exotic $7_3$-branes, respectively.}.
Indeed, explicit solutions of defect $(p,q)$ five-branes are found as
the $SL(2,{\mathbb Z}) \times SL(2,{\mathbb Z})$ conjugate solutions to
a single defect five-brane \cite{Kimura:2014wga}. 
We studied the conjugate configurations of the defect HKT and its T-dualized solution.
The monodromy matrices of these configurations are given by the $O(2,2;{\mathbb Z})$
similarity transformations of $\Omega_{\alpha,\beta}^{\text{dHKT}}$ and $\Omega_{\alpha,\beta}^{\text{TdHKT}}$.
Under the general $O(2,2;{\mathbb Z}) = SL(2,{\mathbb Z}) \times SL(2,{\mathbb Z})$ transformations, 
the explicit solutions are calculated and given by (\ref{ConjugateConfig-dHKT}) and (\ref{ConjugateConfig-TdHKT}).
The conjugate solution (\ref{ConjugateConfig-dHKT}) is interpreted as 
a composite state of $p$ defect NS5-branes and $q$ exotic $5^2_2$-branes
-- a defect $(p,q)$ five-brane -- on the ALG space. 
The solution (\ref{ConjugateConfig-TdHKT}) represents 
other defect $(p,q)$ five-branes on the ALG space.
They are five-brane solutions of codimension two whose higher dimensional origins are hyper-K\"{a}hler geometries with torsion.
Indeed, they are nothing but the local descriptions of the stringy cosmic fivebranes discussed in \cite{Hellerman:2002ax}.

It is interesting to study quantum corrections to the HKT
in the framework of string worldsheet theory.
In order to do that,  it is significant to develop the gauged linear sigma model (GLSM) for exotic five-branes \cite{Kimura:2013fda, Kimura:2013zva, Kimura:2013khz} in much a deeper level.
We can utilize the GLSM to investigate worldsheet instanton corrections to the HKT.
Applying the HKT to the Dirac-Born-Infeld action \cite{Chatzistavrakidis:2013jqa, Kimura:2014upa} for various defect five-branes on the ALG space would be an interesting task to understand non-trivial aspects of exotic five-branes on curved spacetimes.
It would also be important to study physical properties of defect $(p,q)$ five-branes and other composite states of exotic five-branes.
For example, there are various exotic branes whose monodromies are given by the U-duality transformations.
Such branes provide stringy geometries called U-folds.
It might be important for developing the T-duality transformation techniques on GLSMs \cite{Kimura:2014bxa, Kimura:2014aja} to those of the U-duality transformations. 
We will come back to these issues in future studies.

\section*{Acknowledgements}

The authors would like to thank Shun'ya Mizoguchi for helpful discussions. 
The work of TK is supported in part by the Iwanami-Fujukai Foundation.
The work of SS is supported in part by Kitasato University Research Grant for Young Researchers.

\begin{appendix}

\section*{Appendix}

\section{Conventions}
\label{app:convention}

In this appendix, we exhibit the T-duality transformation rules from two different viewpoints as discussed in \cite{Kimura:2014wga}.
In the framework of the background field configuration, 
we mainly utilize the Buscher rule.
Performing T-duality along the $n$-th direction, 
the metric $G_{MN}$, the B-field $B_{MN}$ and the dilaton $\phi$ are transformed in the following way,
\bsubeq \label{Buscher}
\begin{gather}
G'_{MN} \ = \ 
G_{MN} 
- \frac{G_{n M} G_{n N} - B_{n M} B_{n N}}{G_{nn}}
\, , \ls
G'_{n N} \ = \ 
\frac{B_{n N}}{G_{nn}}
\, , \ls
G'_{nn} \ = \ 
\frac{1}{G_{nn}}
\, , \\
B'_{MN} \ = \ 
B_{MN} 
+ \frac{2 G_{n [M} B_{N]n}}{G_{nn}}
\, , \ls
B'_{n N} \ = \ 
\frac{G_{n N}}{G_{nn}}
\, , \\
\phi' \ = \ 
\phi - \half \log (G_{nn})
\, .
\end{gather}
\esubeq
On the other hand, from the perspective of the monodromy transformations,
we use the matrix representations,
\bsubeq \label{T-Umatrix}
\begin{align}
U_8 \ &= \ 
\left(
\begin{array}{cc}
\mathbbm{1} - T_8 & - T_8
\\
- T_8 & \mathbbm{1} - T_8
\end{array}
\right)
\, , \ls
T_8 \ \equiv \ 
\left(
\begin{array}{cc}
1 & 0 
\\
0 & 0
\end{array}
\right)
\, , \label{T8-M} \\
U_9 \ &= \ 
\left(
\begin{array}{cc}
\mathbbm{1} - T_9 & - T_9
\\
- T_9 & \mathbbm{1} - T_9
\end{array}
\right)
\, , \ls
T_9 \ \equiv \ 
\left(
\begin{array}{cc}
0 & 0 
\\
0 & 1
\end{array}
\right)
\, , \label{T9-M} \\
U_{89} \ &= \ 
\left(
\begin{array}{cc}
\mathbbm{1} - T_8 - T_9 & - T_8 - T_9
\\
- T_8 - T_9 & \mathbbm{1} - T_8 - T_9
\end{array}
\right)
\ = \ 
U_8 U_9 \ = \ 
U_9 U_8
\, . \label{T89-M}
\end{align}
\esubeq
These transformation matrices act on an $O(2,2;{\mathbb Z})$ monodromy matrix $\Omega$
as
\begin{align}
\Omega' \ &= \ 
U^{\text{T}} \Omega \, U
\, .
\end{align}

\section{Review of defect five-branes}
\label{app:five-branes}

In this appendix, we briefly summarize the feature of defect five-branes discussed in \cite{Kimura:2014wga}.
In this discussion, we often use the following functions and variables,
\bsubeq
\begin{gather}
H_{\ell} \ = \ 
h + \ell \log \frac{\mu}{\varrho}
\, , \ls
V_{\ell} \ = \ \ell \vartheta
\, , \ls
K_{\ell} \ = \ (H_{\ell})^2 + (V_{\ell})^2
\, , \\
x^6 \ = \ \varrho \cos \vartheta
\, , \ls
x^7 \ = \ \varrho \sin \vartheta
\, , \ls
\ell \ = \ 
\frac{\ell_0}{2 \pi R_8}
\, ,
\end{gather}
\esubeq
where $\ell_0$ and $\ell$ are constant parameters which appear in the main part of this paper as $\ell_{(0)} = \alpha_{(0)}$ or $\ell_{(0)} = \beta_{(0)}$.


First, we discuss a defect NS5-brane and its monodromy structures.
The background configuration is represented as
\bsubeq \label{dNS5-system}
\begin{gather}
\d s^2 \ = \ 
\d s_{012345}^2
+ H_{\ell} \, \big[ (\d \varrho)^2 + \varrho^2 (\d \vartheta)^2 \big]
+ H_{\ell} \, \big[ (\d x^8)^2 + (\d x^9)^2 \big]
\, , \\
B_{89} \ = \ V_{\ell}
\, , \ls
\e^{2 \phi} \ = \ H_{\ell}
\, .
\end{gather}
\esubeq
In terms of this,
the matrix $\mathscr{M}$ and the $O(2,2;{\mathbb Z})$ monodromy matrix $\Omega$ defined in (\ref{MM-MM}) are explicitly formulated as
\begin{align}
\mathscr{M}^{\text{NS}} (\varrho, \vartheta)
\ &= \ 
\frac{1}{H_{\ell}}
\left(
\begin{array}{cccc}
K_{\ell} & 0 & 0 & V_{\ell} \\
0 & K_{\ell} & - V_{\ell} & 0 \\
0 & - V_{\ell} & 1 & 0 \\
V_{\ell} & 0 & 0 & 1
\end{array}
\right)
\, , \ls 
\Omega_{\ell}^{\text{NS}}
\ = \ 
\left(
\begin{array}{cccc}
1 & 0 & 0 & 0 
\\
0 & 1 & 0 & 0 
\\
0 & -2 \pi \ell & 1 & 0
\\
2 \pi \ell & 0 & 0 & 1
\end{array}
\right)
\, . 
\label{SO22-MM-dNS5}
\end{align}
The monodromy matrix denotes that this system is geometric.
We also define the two complex structures associated with the equivalent monodromy group $SL(2,{\mathbb Z})_{\tau} \times SL(2,{\mathbb Z})_{\rho}$.
Their monodromy transformations by the shift $z \to z\,\e^{2 \pi \I}$ give rise to the monodromy matrices,
\bsubeq \label{dNS5-SL2SL2-monodromy}
\begin{alignat}{2}
\tau \ &\to \ 
\tau' \ = \ \tau 
\, , &\ls
\Omega_{\tau}^{\text{NS}}
\ &\equiv \ 
\left(
\begin{array}{cc}
1 & 0 \\
0 & 1
\end{array}
\right)
\, , \\
\rho \ &\to \ 
\rho' \ = \ \rho + 2 \pi \ell 
\, , &\ls
\Omega_{\rho}^{\text{NS}}
\ &\equiv \ 
\left(
\begin{array}{cc}
1 & 2 \pi \ell \\
0 & 1
\end{array}
\right)
\, .
\end{alignat}
\esubeq
This implies that the shape of the fibred two-torus $T^2_{89}$ is not deformed by the monodromy transformation, while the metric and the B-field on $T^2_{89}$ are modified.
We can further discuss the conjugate configuration of the single defect NS5-brane via the transformation (\ref{conjugate-U}),
\begin{align}
\wt{\Omega}_{\tau}^{\text{NS}}
\ &= \ 
\left(
\begin{array}{cc}
1 & 0 
\\
0 & 1
\end{array}
\right)
\, , \ls
\wt{\Omega}_{\rho}^{\text{NS}}
\ = \ 
\left(
\begin{array}{cc}
1 + 2 \pi \ell pq & 2 \pi \ell p^2
\\
- 2 \pi \ell q^2 & 1 - 2 \pi \ell pq
\end{array}
\right)
\, . \label{conjugate-tr-dNS5}
\end{align}
We see that the two-torus $T^2_{89}$ is unchanged under the conjugate.
The matrix $\wt{\Omega}_{\rho}^{\text{NS}}$ implies that the conjugate system is a composite of $p$ defect NS5-branes and $q$ exotic $5^2_2$-branes.


Performing the T-duality transformation along the 9-th direction of (\ref{dNS5-system}),
we obtain the following configuration,
\bsubeq \label{dKK5-system}
\begin{gather}
\d s^2 \ = \ 
\d s_{012345}^2
+ H_{\ell} \, \big[ (\d \varrho)^2 + \varrho^2 (\d \vartheta)^2 \big] 
+ H_{\ell} \, (\d x^8)^2 
+ \frac{1}{H_{\ell}} \big[ \d y^9 - V_{\ell} \, \d x^8 \big]^2 
\, , \\
B_{MN} \ = \ 0
\, , \ls 
\e^{2 \phi} \ = \ 1
\, .
\end{gather}
\esubeq
This is the background geometry of a defect KK5-brane,
where the transverse space of the 6789-directions is an ALG space.
In the main part of this paper, we often refer to this space as the defect Taub-NUT space.
Here the B-field and the dilaton are trivial.
Then the matrices $\mathscr{M}$ and $\Omega$ are simple forms,
\begin{align}
\mathscr{M}^{\text{KK}} (\varrho, \vartheta)
\ &= \ 
\frac{1}{H_{\ell}}
\left(
\begin{array}{cccc}
K_{\ell} & -V_{\ell} & 0 & 0 
\\
-V_{\ell} & 1 & 0 & 0 
\\
0 & 0 & 1 & V_{\ell} 
\\ 
0 & 0 & V_{\ell} & K_{\ell} 
\end{array}
\right)
\, , \ls 
\Omega_{\ell}^{\text{KK}}
\ = \ 
\left(
\begin{array}{cccc}
1 & 0 & 0 & 0 
\\
-2 \pi \ell & 1 & 0 & 0 
\\
0 & 0 & 1 & 2 \pi \ell
\\ 
0 & 0 & 0 & 1
\end{array}
\right)
\, . 
\label{SO22-MM-dKK5}
\end{align}
Under the shift $z \to z \, \e^{2 \pi \I}$,
the $SL(2,{\mathbb Z})_{\tau} \times SL(2,{\mathbb Z})_{\rho}$ monodromy matrices generated by the two complex structures $\tau$ and $\rho$ are given as
\bsubeq \label{dKK5-SL2SL2-monodromy}
\begin{alignat}{2}
\tau \ &\to \ 
\tau' \ = \ \frac{\tau}{- 2 \pi \ell \tau + 1}
\, , &\ls
\Omega_{\tau}^{\text{KK}}
\ &\equiv \ 
\left(
\begin{array}{cc}
1 & 0 \\
- 2 \pi \ell & 1
\end{array}
\right)
\, , \\
\rho \ &\to \ 
\rho' \ = \ \rho 
\, , &\ls
\Omega_{\rho}^{\text{KK}}
\ &\equiv \ 
\left(
\begin{array}{cc}
1 & 0 \\
0 & 1
\end{array}
\right)
\, .
\end{alignat}
\esubeq
These monodromy transformations provide that the field configuration $G_{mn}$ and $B_{89}$ is unchanged, while the shape of the two-torus $T^2_{89}$ is deformed.
We apply the conjugate transformation (\ref{conjugate-U}) to these $SL(2,{\mathbb Z})_{\tau} \times SL(2,{\mathbb Z})_{\rho}$ monodromy matrices,
\begin{align}
\wt{\Omega}_{\tau}^{\text{KK}}
\ &= \ 
\left(
\begin{array}{cc}
1 + 2 \pi \ell r's' & 2 \pi \ell r'{}^2
\\
- 2 \pi \ell s'{}^2 & 1 - 2 \pi \ell r's'
\end{array}
\right)
\, , \ls
\wt{\Omega}_{\rho}^{\text{KK}}
\ = \ 
\left(
\begin{array}{cc}
1 & 0 
\\
0 & 1
\end{array}
\right)
\, . \label{conjugate-tr-dKK5}
\end{align}
Again we see that the field configuration is unchanged but the two-torus $T^2_{89}$ is deformed.
In particular, we find that the conjugate system is the bound state of $-s'$ defect KK5-branes and $r'$ defect KK5-branes of another type \cite{Kimura:2014wga}.


If we take the T-duality transformation along the 8-th direction of (\ref{dNS5-system}), 
we obtain another configuration of the defect KK5-brane of different type,
\bsubeq \label{adKK5-system}
\begin{gather}
\d s^2 \ = \ 
\d s_{012345}^2
+ H_{\ell} \, \big[ (\d \varrho)^2 + \varrho^2 (\d \vartheta)^2 \big]
+ H_{\ell} \, (\d x^9)^2 
+ \frac{1}{H_{\ell}} \big[ \d y^8 + V_{\ell} \, \d x^9 \big]^2 
\, , \\
B_{MN} \ = \ 0
\, , \ls
\e^{2 \phi} \ = \ 1
\, .
\end{gather}
\esubeq
In this setup, the B-field and the dilaton are again trivial.
The $O(2,2;{\mathbb Z})$ monodromy structure (\ref{MM-MM}) is explicitly formulated as
\begin{align}
\mathscr{M}^{\text{AK}} (\varrho, \vartheta)
\ &= \ 
\frac{1}{H_{\ell}}
\left(
\begin{array}{cccc}
1 & V_{\ell} & 0 & 0 
\\
V_{\ell} & K_{\ell} & 0 & 0 
\\
0 & 0 & K_{\ell} & - V_{\ell}
\\
0 & 0 & - V_{\ell} & 1
\end{array}
\right)
\, , \ls 
\Omega_{\ell}^{\text{AK}}
\ = \ 
\left(
\begin{array}{cccc}
1 & 2 \pi \ell & 0 & 0 
\\
0 & 1 & 0 & 0 
\\
0 & 0 & 1 & 0 
\\
0 & 0 & - 2 \pi \ell & 1
\end{array}
\right)
\, . 
\label{SO22-MM-adKK5}
\end{align}
The $SL(2,{\mathbb Z})_{\tau} \times SL(2,{\mathbb Z})_{\rho}$ monodromy structure is also immediately obtained by the definition (\ref{field-config}) and the shift $z \to z \, \e^{2 \pi \I}$ in the 67-plane,
\bsubeq \label{adKK5-SL2SL2-monodromy}
\begin{alignat}{2}
\tau \ &\to \ 
\tau' \ = \ \tau + 2 \pi \ell 
\, , &\ls
\Omega_{\tau}^{\text{AK}}
\ &\equiv \ 
\left(
\begin{array}{cc}
1 & 2 \pi \ell \\
0 & 1
\end{array}
\right)
\, , \\
\rho \ &\to \ 
\rho' \ = \ \rho 
\, , &\ls
\Omega_{\rho}^{\text{AK}}
\ &\equiv \ 
\left(
\begin{array}{cc}
1 & 0 \\
0 & 1
\end{array}
\right)
\, .
\end{alignat}
\esubeq
This also indicates that the two-torus $T^2_{89}$ is deformed under the monodromy transformation. 
This property is the same as in (\ref{dKK5-system}).
According to the transformations (\ref{conjugate-U}),
we obtain the conjugate monodromy matrices,
\begin{align}
\wt{\Omega}_{\tau}^{\text{AK}}
\ &= \ 
\left(
\begin{array}{cc}
1 + 2 \pi \ell p'q' & 2 \pi \ell p'{}^2
\\
-2 \pi \ell q'{}^2 & 1 - 2 \pi \ell p'q'
\end{array}
\right)
\, , \ls
\wt{\Omega}_{\rho}^{\text{AK}}
\ = \ 
\left(
\begin{array}{cc}
1 & 0
\\
0 & 1
\end{array}
\right)
\, . \label{conjugate-tr-adKK5}
\end{align}
This leads to a composite of $q'$ defect KK5-branes and $p'$ defect KK5-branes of another type.


The final example is the background geometry of an exotic $5^2_2$-brane.
Its spacetime configuration is given by
\bsubeq \label{522-system}
\begin{gather}
\d s^2 \ = \ 
\d s_{012345}^2
+ H_{\ell} \, \big[ (\d \varrho)^2 + \varrho^2 (\d \vartheta)^2 \big]
+ \frac{H_{\ell}}{K_{\ell}} \big[ (\d y^8)^2 + (\d y^9)^2 \big]
\, , \\
B_{89} \ = \ - \frac{V_{\ell}}{K_{\ell}}
\, , \ls
\e^{2 \phi} \ = \ \frac{H_{\ell}}{K_{\ell}}
\, .
\end{gather}
\esubeq
This gives rise to the matrices of the $O(2,2;{\mathbb Z})$ monodromy structure,
\begin{align}
\mathscr{M}^{\text{E}} (\varrho, \vartheta)
\ &= \ 
\frac{1}{H_{\ell}}
\left(
\begin{array}{cccc}
1 & 0 & 0 & - V_{\ell} 
\\
0 & 1 & V_{\ell} & 0 
\\
0 & V_{\ell} & K_{\ell} & 0 
\\
- V_{\ell} & 0 & 0 & K_{\ell}
\end{array}
\right)
\, , \ls 
\Omega_{\ell}^{\text{E}}
\ = \ 
\left(
\begin{array}{cccc}
1 & 0 & 0 & - 2 \pi \ell
\\
0 & 1 & 2 \pi \ell & 0 
\\
0 & 0 & 1 & 0 
\\
0 & 0 & 0 & 1
\end{array}
\right)
\, . 
\label{SO22-MM-522}
\end{align}
We can confirm that this configuration is nongeometric because the monodromy matrix $\Omega_{\ell}^{\text{E}}$ involves the $2 \times 2$ block $\beta$ of (\ref{ADTB-O22}).
We also mention that the structure of the $SL(2,{\mathbb Z})_{\tau} \times SL(2,{\mathbb Z})_{\rho}$ monodromy is dictated by the two complex structures $\tau$ and $\rho$.
They are transformed when we go around the exotic $5^2_2$-brane with $z \to z \e^{2 \pi \I}$,
\bsubeq \label{522-SL2SL2-monodromy}
\begin{alignat}{2}
\tau \ &\to \ 
\tau' \ = \ \tau 
\, , &\ls
\Omega_{\tau}^{\text{E}}
\ &\equiv \ 
\left(
\begin{array}{cc}
1 & 0 \\
0 & 1
\end{array}
\right)
\, , \\
\rho \ &\to \ 
\rho' \ = \ \frac{\rho}{- 2 \pi \ell \rho + 1}
\, , &\ls
\Omega_{\rho}^{\text{E}}
\ &\equiv \ 
\left(
\begin{array}{cc}
1 & 0 \\
- 2 \pi \ell & 1
\end{array}
\right)
\, .
\end{alignat}
\esubeq
This indicates that the two-torus $T^2_{89}$ is invariant under the monodromy transformation, while the field configuration is modified.
The conjugate monodromy matrices defined by the transformation rules (\ref{conjugate-U}) are given as
\begin{align}
\wt{\Omega}_{\tau}^{\text{E}}
\ &= \ 
\left(
\begin{array}{cc}
1 & 0 
\\
0 & 1
\end{array}
\right)
\, , \ls
\wt{\Omega}_{\rho}^{\text{E}}
\ = \ 
\left(
\begin{array}{cc}
1 + 2 \pi \ell rs & 2 \pi \ell r^2
\\
-2 \pi \ell s^2 & 1 - 2 \pi \ell rs
\end{array}
\right)
\, . \label{conjugate-tr-522}
\end{align}
Indeed the conjugate matrices imply that the conjugate system is a composite of $r$ defect NS5-branes and $-s$ exotic $5^2_2$-branes.

\end{appendix}

}
\end{document}